\documentstyle[aps,graphicx]{revtex}
\begin{document}
\draft
\begin{title}
{Anisotropic Magnetoresistance in Ga$_{1-x}$Mn$_x$As }

\end{title}
\author{David V. Baxter, Dmitry Ruzmetov, Julia Scherschligt}
\address{Dept. of Physics, Indiana University, Bloomington Indiana}
\author{Y. Sasaki, X. Liu, J. K. Furdyna}
\address{Department of Physics, University of Notre Dame, Notre Dame,
IN}
\author{C. H. Mielke}
\address{National High Magnetic Field Laboratory, Los Alamos National
Lab, New Mexico}
\date{\today}
 \twocolumn[\hsize\textwidth\columnwidth\hsize\csname
 @twocolumnfalse\endcsname
 \maketitle
\begin{abstract}
We have measured the magnetoresistance in a series of Ga$_{1-x}$Mn$_x$As
samples with 0.033$\le x \le$ 0.053 for three mutually orthogonal
orientations of the applied magnetic field.  The spontaneous resistivity
anisotropy (SRA) in these materials is negative (i.e. the sample resistance is
higher when its magnetization is perpendicular to the measuring
current than when the two are parallel) and
has a magnitude on the order of 5\% at temperatures near 10K and below.
This stands in contrast to the results for most conventional magnetic
materials where the SRA is considerably smaller in magnitude for
those few cases in which a negative sign is observed.  The magnitude
of the SRA drops from its maximum at low temperatures to zero at T$_C$
in a manner that is consistent with mean field theory.  These results
should provide a significant test for emerging theories of transport
in this new class of materials.
\end{abstract}
 \vskip2pc]


The recent discovery of ferromagnetism at temperatures as high as
110K in Ga$_{1-x}$Mn$_x$As has greatly broadened the interest in
diluted magnetic semiconductors over the past few
years\cite{Ohno96,Ohno98,Omiya00}.  A large number of groups are
now investigating these materials which could form the basis for
a wide variety of new magneto-electronic devices that may be
grown pseudo-morphically on GaAs.
The utility of similar devices has already been
established using nano-composite structures based on
metallic magnetic materials (the so-called giant magnetoresistance
and tunnel junction magnetoresistance\cite{Levy94,Moodera99}).
The semiconducting materials open up new opportunities as they also
introduce the prospects of optical or electronic control of the
magnetic properties\cite{Koshihara97,Ohno00}.

These new materials present many interesting
challenges as we embark on efforts to make use of their
properties in real devices.  These challenges come from
both the fundamental point of view (trying to understand
their properties and what factors control them) and from
the desire to manufacture high quality materials in the face
of severe constraints imposed by the possibility of nucleating
unwanted phases during growth\cite{Ohno98}.
These two challenges are intimately connected since the
density of carriers is expected to influence such material
properties as the Curie temperature and magnetic
anisotropy\cite{Konig00,Abolfath01},
but this density is strongly influenced by defects included
in the structure as a result of the constraints on the
growth conditions.

In view of our limited understanding of these materials it
is prudent to spend some effort exploring the transport
properties of individual material layers in preparation for
constructing multilayered structures to form devices.
Toward that end, in this article we report on magnetotransport
measurements in a series of
Ga$_{1-x}$Mn$_x$As films ($x$= 0.033 to 0.053).
We concentrate on the dependence of the film
resistivity on the orientation of an applied
magnetic field in the hope that this
dependence might be less sensitive to the
details of the disorder in the films than is the resistivity itself.

The Mn-alloyed films were grown at temperatures in the range of
275 to 285$^\circ$C.  The substrates were cleaved from GaAs
wafers, and prior to growth of the alloy two GaAs buffer layers
were deposited (the first one 100 nm thick and grown at
550$^\circ$C, and the second one 300nm thick and grown at the
alloy growth temperature). The important physical properties of
the films are listed in table I.  The nominal concentrations were
determined by comparing the film lattice constant (determined
from the 004 reflection) to the values given by Ohno et
al.\cite{Ohno96} (but note that this can lead to systematic
errors as the lattice constant can also depend on the growth
conditions as pointed out by Schott et al.\cite{Schott01}).  The
paramagnetic Curie point for each film was determined by assuming
that the Hall resistance is dominated by the extraordinary Hall
effect for small fields and temperatures not too far above
T$_C$.  We then linearly extrapolate the inverse low-field Hall
resistance above T$_C$ to zero in order to establish T$_C$. The
films were patterned into standard Hall-bar geometry (sample
dimensions were 3.4 mm long with 1.8 mm between the longitudinal
voltage leads and 350 $\mu$m wide) using conventional
photolithography and etching in a solution of
(H$_{2}$SO$_{4}$:H$_{2}$O$_{2}$:H$_{2}$O) in the ratio (1:8:10).
The resistance was measured using ohmic In contacts to connect the
sample to a Quantum Design 1802 bridge. Fields up to 6T were
applied over a temperature range extending from 1.4K to above
T$_C$.  The slight compressive strain imposed on the films by the
substrate results in an intrinsic magnetic anisotropy with an
in-plane easy axis\cite{Ohno98,Abolfath01}.

\begin{figure}
\centering
\includegraphics[width=3in]{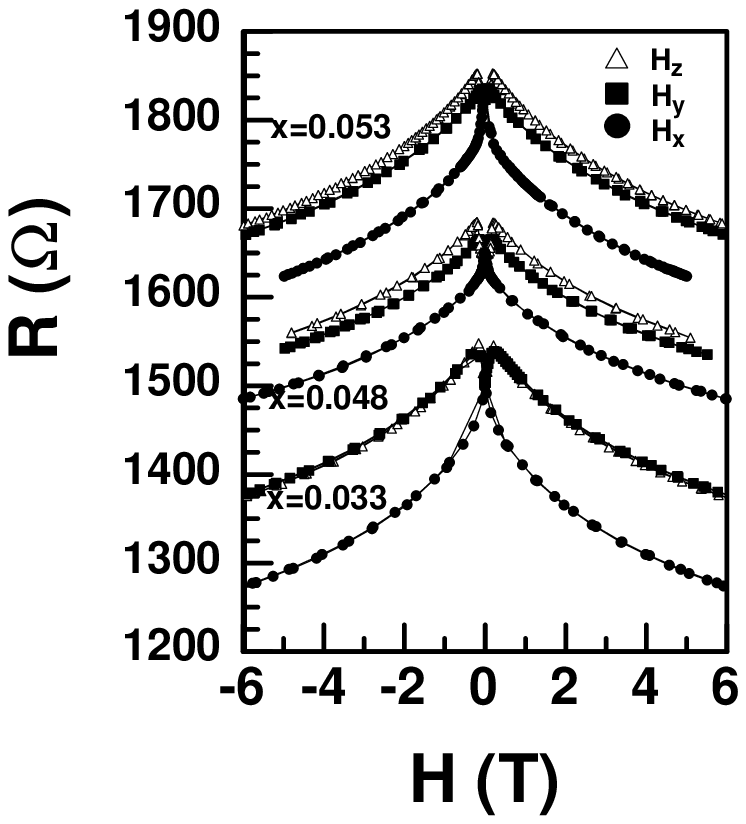}
 \caption{Resistance as a function of magnetic field at
$T = 15{\rm K}$ for Ga$_{1-x}$Mn$_{x}$As films with three
different values of x and three different orientations of the
magnetic field.} \label{Fig1}
\end{figure}

In describing this orientational dependence of the
magnetotransport we will use a coordinate system in which the
z-axis defines the film growth direction and the x-axis
lies along the direction of the measuring current.
Figure~\ref{Fig1} shows the magnetoresistance for three
Ga$_{1-x}$Mn$_x$As films (x=0.033, 0.048, and 0.053) for
situations where the magnetic field is oriented along each of the
three axes at a temperature of 15K (here and in the following we
use the notation H$_x$, H$_y$, and H$_z$ to refer to an applied
field directed along the indicated axis).  Similar data are seen
for other temperatures below the Curie Temperature. Two features
are immediately apparent from the data in this figure. First,
there is the overall negative magnetoresistance seen for large
fields, and second, the resistance obtained by
extrapolating the high-field behavior to zero field differs for
the three field orientations.  The first has been noted before,
and it has been attributed to a field-induced reduction in the
spin disorder responsible for much of the resistance near T$_C$
\cite{Matsukura98}. The second feature is the spontaneous
resistivity anisotropy (SRA), which has not yet been systematically
studied in the new ferromagnetic semiconductors but is a familiar
magnetotransport property in conventional magnetic
materials\cite{Jaoul77}.

\begin{figure}
\centering
\includegraphics[width=3in]{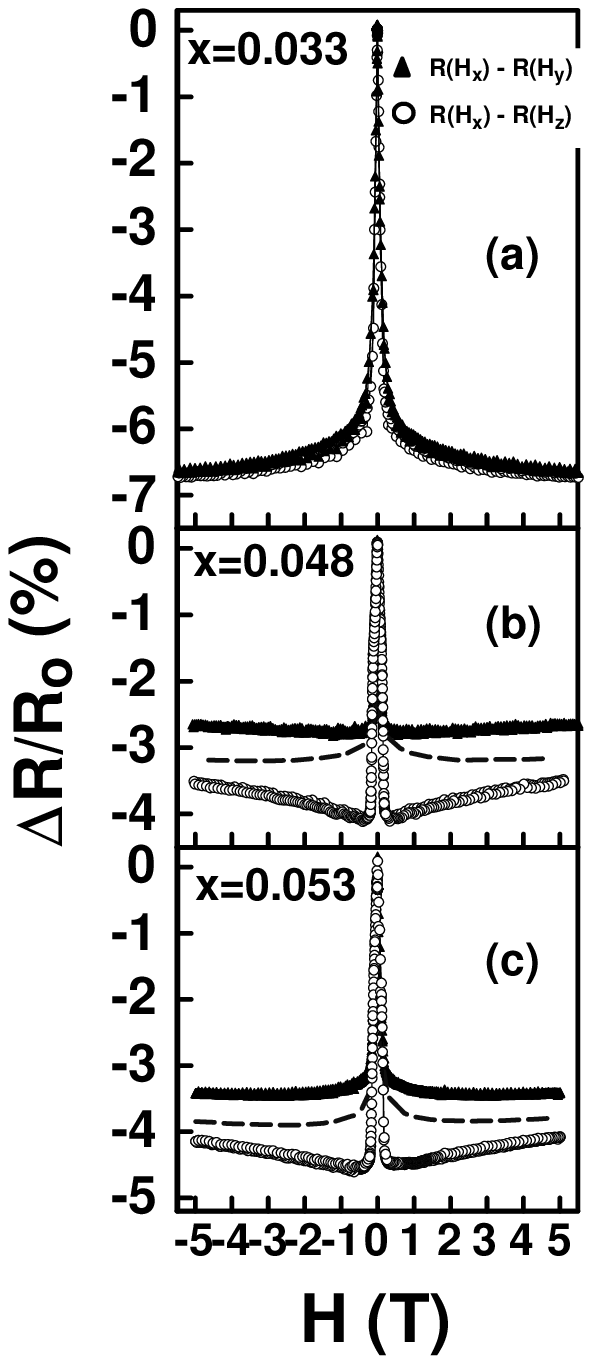}
\caption{Anisotropic magnetoresistance obtained from the data in
fig. 1 by subtracting from the data obtained with fields applied along the
direction of the current (the x-direction) those date taken with
the field oriented along the y or z axes (giving AMR$_\parallel$
and AMR$_\perp$ respectively). Ignoring small effects
from demagnetization, the extrapolation of the high-field value
of this difference to zero field defines the Spontaneous Resistivity
Anisotropy for both the in-plane and out-of-plane conditions. The dashed
line shown in b and c demonstrates that the upward trend seen in
AMR$_\perp$ for these two samples may be attributed to an
extra ordering field that suppresses in-plane spin disorder when
the magnetization lies in the plane of the sample (see text).}
\label{Fig2}
\end{figure}

Figure~\ref{Fig2} exhibits the anisotropic magnetoresistance (AMR) as
differences between the magnetoresistance (MR) seen for fields applied
perpendicular to the current (i.e. H$_z$ or H$_y$) and that seen for
a field applied along the current (H$_x$). In this figure these
differences have been normalized to the zero-field value of the
resistance. The rapid drop in the AMR seen at low fields in this
figure represents the SRA and we note that, contrary to the
behavior seen in most conventional ferromagnets,
the resistivity of these new materials is greater when the
magnetization is oriented perpendicular to the current than when
it lies parallel to the current.  This no doubt contributes to the
familiar low-field peaks seen in typical MR data on
Ga$_{1-x}$Mn$_x$As for fields oriented perpendicular to the sample
plane\cite{Matsukura98}. When the applied field is sufficient
to overcome the intrinsic anisotropy of the material,
the resistivity increases as some domains
that were originally aligned with the current are reoriented
out-of-the-plane (thereby increasing the resistivity of that
portion of the sample).  To confirm this interpretation we note that
the peaks are also present when the field is applied in the plane
so long as it is perpendicular to the current, but are absent
when the field is applied along the current.

Looking at Fig.~\ref{Fig2}, it is apparent that the AMR for the
two more concentrated samples is not restricted to low fields.
The out-of-plane AMR (AMR$_\perp$) shows a distinct field dependence
above 0.5T for x=0.048 and 0.053, which indicates that for these samples
the high-field MR also depends on the field orientation. As-grown samples
of GaMnAs exhibit a distinct peak in the temperature dependence of
their resistivity and this peak has been demonstrated to arise from
spin-fluctuations by comparing the temperature variation of the
resistivity above T$_C$ to the magnetic susceptibility (as expected
from the fluctuation-dissipation theorem)\cite{Ohno99b,Dietl94}. If we
assume that high-field MR is due to the suppression of similar
spin disorder by the applied field, then the behavior exhibited in
fig.~\ref{Fig2} indicates an anisotropy in that spin disorder below
T$_C$.  To gain some insight into the origin of this anisotropy we
consider the dashed lines shown in figures 2b and 2c, which
represent the quantity (R(H$_z$+H$_o$) -R(H$_x$)), as
a function of the applied field strength.  Here, the
field strength is given by $|H_z|$=$|H_x|$, and H$_o$
represents a small offset field (0.4T for x=0.048 and 0.3T for
x=0.053).  This ``shifted-field'' AMR$_\perp$
data for these two compositions looks remarkably like the in-plane
AMR data in character.  We therefore attribute the upward trend
seen in the raw AMR$_\perp$ data (open circles in
Fig.~\ref{Fig2}) to an extra in-plane ordering field.  We note, however,
that this extra field is not sufficient to account entirely for the observed
difference between the in-plane and out-of-plane SRA (SRA$_\parallel$ and
SRA$_\perp$ respectively below) in our two more
concentrated samples.  This suggests that the SRA is not dependent solely
on the relative orientation of the current and the magnetization in these
two samples; there is also a contribution that depends on the orientation
of these two vectors with respect to the sample's crystal structure which
has a slight tetragonal distortion due to the substrate.

\begin{figure}
\centering
\includegraphics[width=3in]{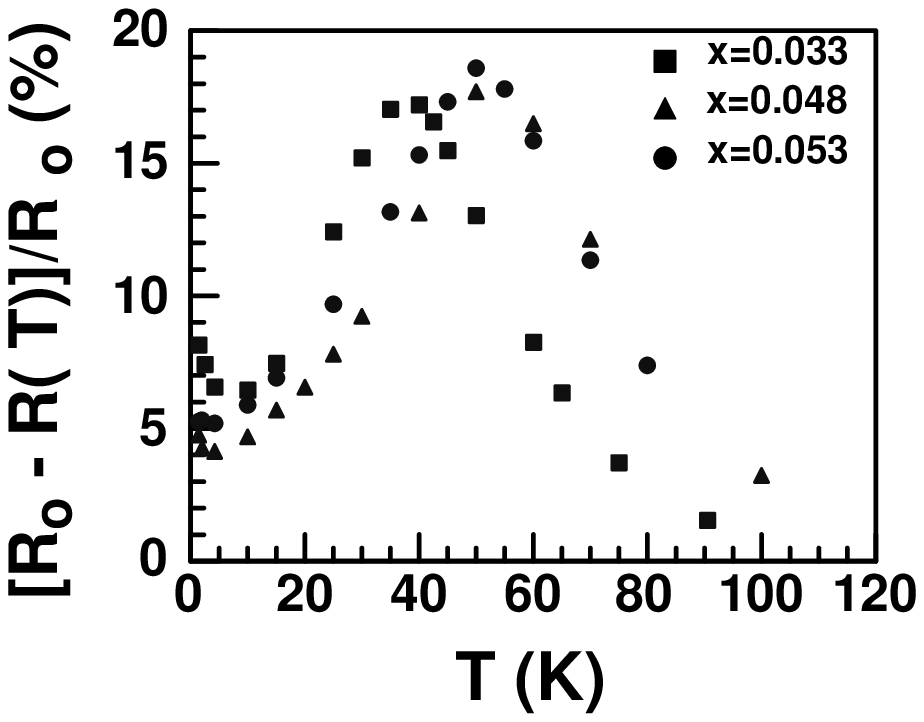}
\caption{Temperature dependence of the magnitude of the longitudinal
magnetoresistance at relatively high fields.  We define the
MR as the resistivity in a field of 5T applied parallel to the
current minus the resistivity in zero field normalized to the
latter value.  Note the peak near T$_C$ and the non-zero value
reached for low temperatures.} \label{Fig3}
\end{figure}

Some additional insight into the disorder present in these films
may be obtained by looking at the temperature dependence of the
high-field MR.  In fig.~\ref{Fig3} we plot the MR (defined as the
difference between the zero-field value of the resistance and the
resistance at 5T, normalized to the zero-field value) as a
function of temperature.  This MR reaches a maximum magnitude near
T$_C$.  This helps to confirm magnetic disorder as the origin
of the well-known peak in the resistivity itself near T$_C$ which has
been attributed to critical spin-fluctuations\cite{Ohno99b}.
Interestingly, the MR does not
approach zero as the temperature is lowered but rather saturates
at a value near -5\%. This suggests that a substantial portion of
the spin disorder in these samples is not critical, or even thermal, in
origin but rather reflects the presence of frustration in the spin system.
The data shown in this figure are taken with the field aligned
parallel to the measuring current, but similar data (though they
differ slightly in magnitude) are seen for the other two
orientations of the magnetic field.

Spin-orbit coupling lies at the heart of any description of the SRA (since
the phenomenon itself intimately links the carrier spin with its orbital
motion) but quantitative descriptions also depend upon the details of the Fermi
surface and a model of transport for the materials in question.
Theoretical descriptions of the SRA
in conventional magnetic materials are typically described within a
parallel conduction model for the transport (one channel for spin-up electrons,
another for spin-down electrons) and spin-orbit coupling is treated as
a perturbation \cite{Jaoul77,Malozemoff85}.  However, in Ga$_{1-x}$Mn$_x$As
the spin-orbit coupling strength is comparable to the Fermi
Energy\cite{Abolfath01} so such a parallel conduction model for transport
is not appropriate for these materials.
In order to assist those trying to develop more appropriate theories for these
materials, we provide some additional information on our samples along with some
qualitative analysis below.

\begin{figure}
\centering
\includegraphics[width=3in]{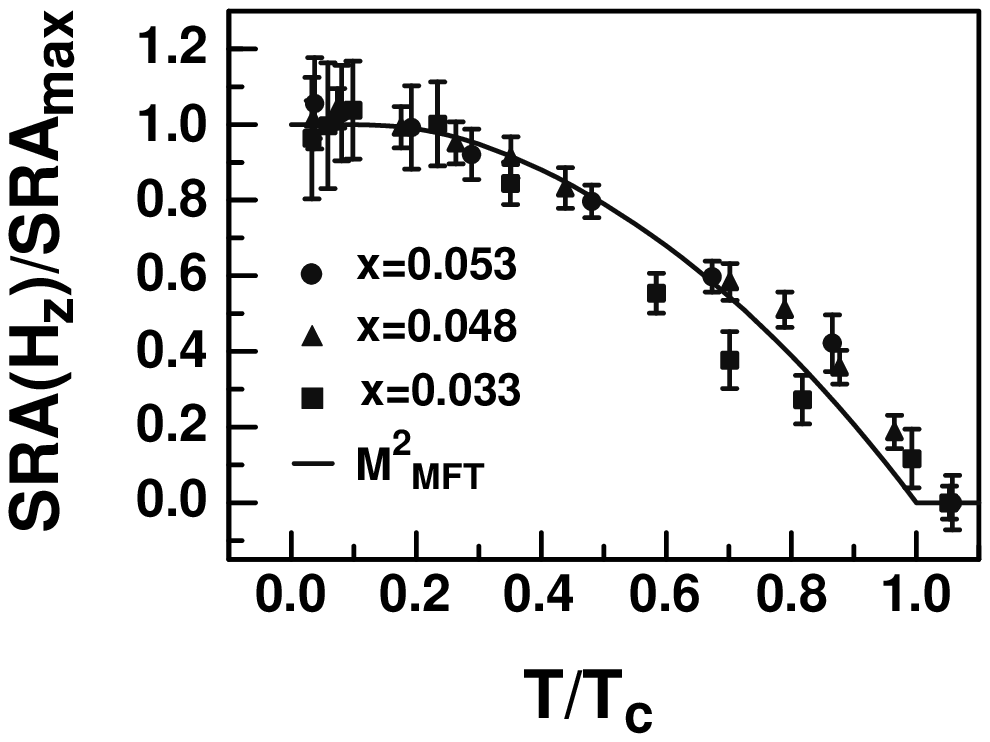}
\caption{The normalized temperature dependence of the SRA$_\perp$
for the three samples compared to the variation expected for the
square of the magnetization in a mean-field Heisenberg
ferromagnet with S=5/2.} \label{Fig4}
\end{figure}

We have determined the carrier concentration in our samples
by measuring the Hall resistance in very large magnetic fields (up to 55T)
at low temperatures (down to 600mK) using the pulsed-field facility
at Los Alamos National Lab \cite{Scherschligt01}.
For such extreme conditions, the extraordinary effect should saturate
and therefore the slope of Hall resistance vs. applied field for fields above
20T can be used to determine
the carrier concentration (p) \cite{Omiya00}.  The carrier concentrations
so derived are displayed in table I.  We also note, as displayed in
fig.~\ref{Fig4}, that the temperature dependence of SRA$_\perp$ is consistent
with the behavior expected for a mean-field Heisenberg model.  This
indicates that a mean-field approach might be adequate to describe
the SRA even though its validity for predicting the Curie temperature
and the temperature dependence of the magnetization has been called
into question\cite{Konig00,Chattopadhyay01b,Berciu01}.  The adequacy of
the mean-field approach in this case may reflect our extrapolation from
field greater than 0.3T back to zero in determining the SRA.
Fields this large may be sufficient to suppress any fluctuations in these
samples beyond those expected from mean-field behavior.

\acknowledgements

The authors gratefully acknowledge discussions with A. H. MacDonald
and M. Abolfath.  This work was supported by the Office of Naval Research
and the Research Foundation fo the State University of New York
under grant number N000140010951 and by the 21st Century Science
and Technology Fund of the State of Indiana.


\begin{thebibliography}{10}

\bibitem{Ohno96}
H. Ohno {\it et~al.}, Applied Physics Letters {\bf 69},  363
(1996).

\bibitem{Ohno98}
H. Ohno, Science {\bf 281},  951  (1998).

\bibitem{Omiya00}
T. Omiya {\it et~al.}, Physica E {\bf 7},  976  (2000).

\bibitem{Levy94}
P.~M. Levy, Solid State Physics {\bf 47},  397  (1994).

\bibitem{Moodera99}
J.~S. Moodera and J. Nassar, Annual Review of Materials Science
{\bf 29},  381
  (1999).

\bibitem{Koshihara97}
S. Koshihara {\it et~al.}, Physical Review Letters {\bf 78},
4617  (1997).

\bibitem{Ohno00}
H. Ohno {\it et~al.}, Nature {\bf 408},  944  (2000).

\bibitem{Konig00}
J. K\"onig, H.-H. Lin, and A.~H. MacDonald, Physical Review
Letters {\bf 84},
  5628  (2000).

\bibitem{Abolfath01}
M. Abolfath, T. Jungwirth, J. Brum, and A.~H. MacDonald, Physical
Review B {\bf
  63},  054418: 14 pages  (2001).

\bibitem{Schott01}
G.~M. Schott, W. Faschinger, and L.~W. Molenkamp, Applied Physics
Letters {\bf
  79},  1807  (2001).

\bibitem{Matsukura98}
F. Matsukura, H. Ohno, A. Shen, and Y. Sugawara, Physical Review
B {\bf 57},
  R2037  (1998).

\bibitem{Jaoul77}
O. Jaoul, I.~A. Campbell, and A. Fert, Journal of Magnetism and
Magnetic
  Materials {\bf 5},  23  (1977).

\bibitem{Ohno99b}
H. Ohno, Journal of Magnetism and Magnetic Materials {\bf 200},
110  (1999).

\bibitem{Dietl94}
T. Dietl,  in {\em Handbook on Semiconductors}, edited by T.~S.
Moss and S.
  Mahajan (Elsevier Science, New York, 1994), pp.\ 1251--1342.

\bibitem{Malozemoff85}
A.~P. Malozemoff, Physical Review B {\bf 32},  6080  (1985).

\bibitem{Scherschligt01}
J. Scherschligt {\it et~al.}, unpublished.

\bibitem{Chattopadhyay01b}
A. Chattopadhyay, S. Das~Sarma, and A.~J. Millis, Physical Review
Letters {\bf
  87},  227202  (2001).

\bibitem{Berciu01}
M. Berciu and R.~N. Bhatt, Physical Review Letters {\bf 87},
107203  (2001).

\end{thebibliography}

%
%

\newpage

%
%

\begin{table}
\caption{\label{table1} Physical Properties of Ga$_{1-x}$Mn$_x$As
Films }
\begin{tabular}{cccccccc}
x  & t  & T$_G$  &
$\rho_{300}$  & $\rho_{4.2}$  &  $T_C$ & p & SRA$_\perp$ \\
(nom.) &  (nm) &  ($^\circ$C) &
($\mu\Omega m$) & ($\mu\Omega m$) &  (K)& nm$^{-3}$ & (at 4.2K)\\
$\pm$ 0.005& $\pm$30nm & $\pm$ 5 $^\circ$C &  $\pm$5\% &
$\pm$5\% & $\pm$3K & $\pm 20\%$ & $\pm 10\%$\\
\tableline
0.033 &  300  &  285  &  54  & 67 &  43 & 0.37 & $-0.070$\\
0.048 &  300  &  275  &  57  & 71 &  57 & 0.30 & $-0.045$ \\
0.053 &  300  &  275  &  61  & 79 &  52 & 0.23 & $-0.051$\\
\end{tabular}
\end{table}

\end{document}